\documentclass[aps,prb,reprint,amsmath,amssymb,floatfix,superscriptaddress]{revtex4-2}
\usepackage{graphicx}
\usepackage{bm}
\usepackage{dcolumn}
\usepackage{booktabs}
\usepackage{hyperref}
\usepackage{xcolor}
\definecolor{journalcol}{RGB}{0,83,159}
\usepackage{etoolbox}
\pretocmd{\bibitem}{\vspace{8pt}}{}{}
\usepackage{placeins}
\usepackage[export]{adjustbox} 
\hypersetup{
    colorlinks=true,
    linkcolor=red,
    filecolor=blue,      
    urlcolor=blue,
    citecolor=blue,
}
\newcommand{\subred}[2]{\textcolor{red}{\ref{#1}(#2)}}
\usepackage{etoolbox}

\AtBeginEnvironment{thebibliography}{\interlinepenalty=10000}

\begin{document}

\title{Nonlinear Hall effect in Floquet-driven monolayer 1T$'$-MoS$_2$}

\author{Muhammad Faisal}
\affiliation{Department of Physics, Quaid-I-Azam University, Islamabad 45320, Pakistan}

\author{Muzamil Shah}
\email{muzamil@qau.edu.pk}
\affiliation{Department of Physics, Quaid-I-Azam University, Islamabad 45320, Pakistan}
\affiliation{Research Center of Astrophysics and Cosmology, Khazar University, 41 Mehseti Street, Baku AZ1096, Azerbaijan}

\author{Imtiaz Khan}
\email{ikhanphys1993@gmail.com}
\affiliation{Department of Physics, Zhejiang Normal University, Jinhua 321004, PR China}
\affiliation{Zhejiang Institute of Photoelectronics \& Zhejiang Institute for Advanced Light Source, Zhejiang Normal University, Jinhua 321004, PR China}

\author{Reza Asgari}
\email{asgari@theory.ipm.ac.ir}
\affiliation{School of Physics, Institute for Research in Fundamental Sciences (IPM), Tehran 19395-5531, Iran}
\date{\today}

\begin{abstract}
\smallskip
\smallskip

We study the nonlinear Hall effect in Floquet-driven monolayer \(1T'\)-MoS\(_2\), a low-symmetry quantum spin Hall material whose tilted Dirac bands sustain an intrinsic Berry-curvature dipole without the need for strain or trigonal warping. We show that off-resonant circularly polarized light offers a way to control both the sign and the magnitude of the nonlinear Hall response through optically induced topological phase transitions using a Floquet effective Hamiltonian and nonlinear semiclassical transport theory. We show that the anisotropic crystal symmetry enforces a selection rule in which the Berry-curvature dipole elements satisfy $D_x\equiv0$, while a finite $D_y$ originates from the intrinsic band tilt. The Berry curvature is recreated in momentum space as the Floquet drive successively inverts individual spin-valley sectors, resulting in an identical sign reversal of the nonlinear Hall conductivity and the Berry-curvature dipole at each bulk gap closing. In contrast, tuning the band tilt modifies only the magnitude of the response without changing its sign, establishing the observed sign reversal as an unambiguous transport signature of genuine Floquet topological phase transitions. We further show that the nonlinear Hall response can be controlled by the driving strength, perpendicular electric field, Fermi energy, and temperature, providing multiple experimental knobs for observation. Our findings establish the sign of the nonlinear Hall response as a universal transport fingerprint of Floquet-engineered topology and point to monolayer \(1T'\)-MoS\(_2\) as a viable platform for all-electrical detection of nonequilibrium topological phases.

\end{abstract}

\maketitle

\section{Introduction}\label{sec:intro}
The nonlinear Hall effect has emerged as a powerful probe of band geometry in time-reversal-symmetric but inversion-broken systems, where a transverse second-order current arises in the absence of an external magnetic field~\cite{sodemann2015,ortix2021,bandyopadhyay2024non,Tang_2017,ahmad2026lengthvelocitygaugeequivalencequantum}. Its central quantity is the Berry-curvature dipole, the first momentum-space moment of the Berry curvature over the occupied states, which is highly sensitive to the symmetry and geometry of electronic bands near the Fermi level~\cite{sodemann2015,xiao2010,Jiang_2025}. Since its theoretical prediction~\cite{sodemann2015}, the nonlinear Hall effect has been experimentally observed in WTe$_2$~\cite{ma2019observation,kang2019}, Bi$_2$Se$_3$-family topological insulators~\cite{he2021}, and strained graphene~\cite{battilomo2019,Du_2021}, firmly establishing broken inversion symmetry as the essential prerequisite for a finite Berry-curvature dipole. In all these systems, however, the Berry-curvature dipole originates from static structural asymmetries, such as trigonal warping or strain~\cite{C7RA10304B,saynatjoki2017ultra,sun2025giant,rose2013spin}, and although its magnitude can be tuned by electrostatic gating or chemical potential, dynamic optical control of its sign and symmetry has remained largely unexplored.

\begin{figure}[htbp]
\centering
\includegraphics[width=\columnwidth]{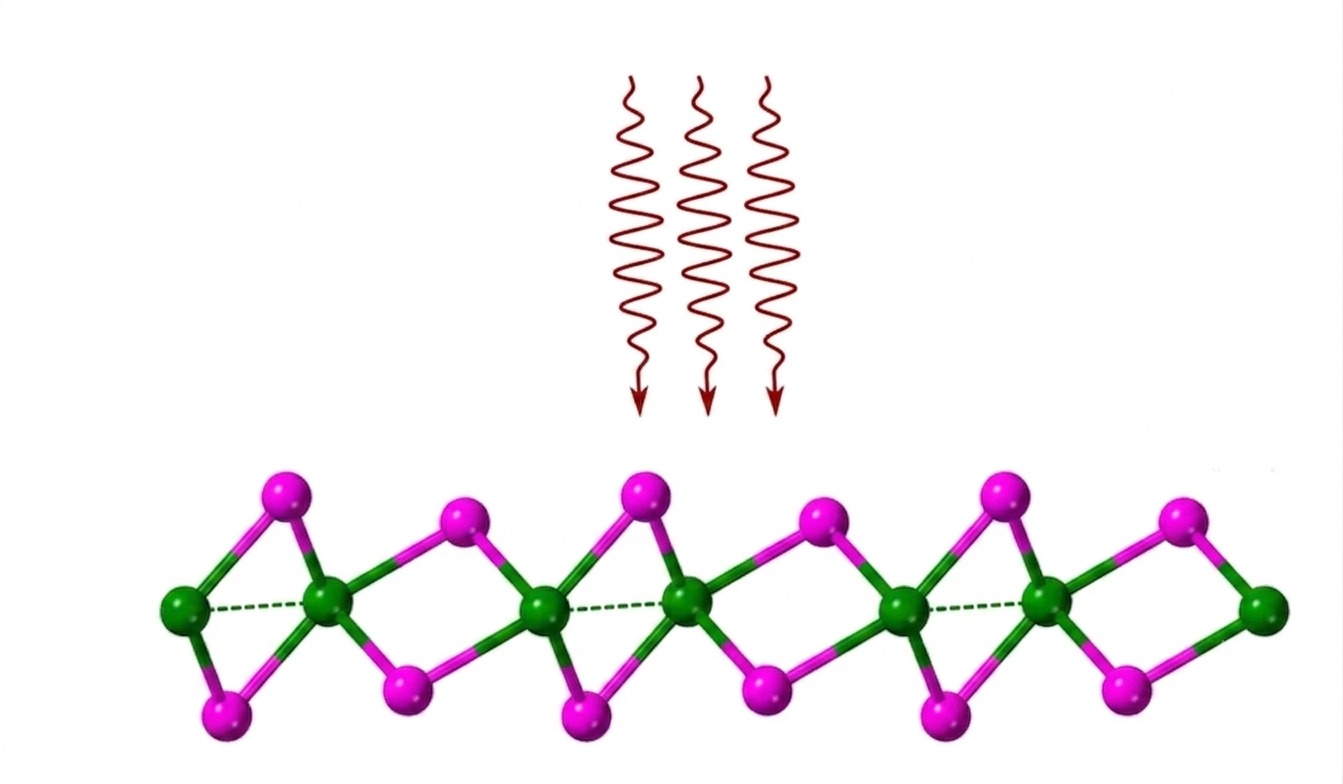}
\caption{Side-view schematic of monolayer \(1T'\)-MoS$_2$, showing Mo atoms (green) and S atoms (magenta), illuminated by off-resonant circularly polarized light. The intrinsic tilt structure induces asymmetry of the electronic bands.}
\label{cartoon}
\end{figure}

Among two-dimensional quantum materials, monolayer \(1T'\)-MoS$_2$ occupies a unique position owing to its distorted crystal structure, strong spin-orbit coupling, and inverted electronic bands~\cite{qv3b-ntkc,qian2014}. As illustrated in Fig.~\ref{cartoon}, the combination of lattice distortion and spin-orbit interaction opens a sizable topological gap and realizes the quantum spin Hall phase~\cite{tang2017,fei2017,wu2018,mortezaei2025}. Unlike the high-symmetry 2H transition-metal dichalcogenides, the \(1T'\) phase possesses pronounced in-plane anisotropy that breaks rotational symmetry while preserving time-reversal symmetry, thereby allowing a finite Berry-curvature dipole without requiring an external magnetic field. Furthermore, its small inverted band gap places the system close to a topological phase boundary, rendering the Berry curvature exceptionally sensitive to external perturbations such as off-resonant circularly polarized light. Consequently, Floquet driving can efficiently renormalize the effective Dirac masses, induce band inversions, and trigger topological phase transitions that are directly reflected in the nonlinear Hall response. Compared with the extensively studied \(1T'\)-WTe$_2$, monolayer \(1T'\)-MoS$_2$ provides a physical theoretical platform with weaker intrinsic disorder, greater tunability through strain, gating, and optical fields, and excellent compatibility with van der Waals heterostructures. These characteristics make it an ideal system for exploring the interplay between Floquet topology, Berry-curvature dipoles, and nonlinear quantum transport, while offering physical prospects for experimental verification.

The low-energy electronic structure of monolayer \(1T'\)-MoS$_2$ is well described by an anisotropic tilted two-band $\mathbf{k}\cdot\mathbf{p}$ Hamiltonian~\cite{das2020,wangD2023,Zhou_2026}. Its tilted Dirac cones are essentially different from the trigonally warped bands found in conventional 2H-phase transition-metal dichalcogenides~\cite{dai2023recent}. This distinction has important consequences for nonlinear transport. Whereas the Berry-curvature dipole in trigonal systems is generated primarily by Fermi-surface warping~\cite{battilomo2019}, the tilted electronic dispersion of \(1T'\)-MoS$_2$~\cite{Balassis2022Polarizability} produces a finite dipole through a distinct mechanism arising from intrinsic band anisotropy. Whether this intrinsic Berry-curvature dipole can be reconstructed by optical driving and whether such reconstruction yields measurable transport signatures has not yet been investigated.

There are very few empirically accessible electrical signs of nonequilibrium topological phases, despite great advancements in Floquet engineering of topological materials. The majority of research has used edge-state detection or spectroscopic measures to discover Floquet-induced band inversions. The nonlinear Hall effect, which is controlled by the Berry-curvature dipole, offers a promising bulk transport probe of Floquet transitions since it is extremely sensitive to modifications in band topology. A feasible method for electrically identifying light-induced topological states would be to establish a direct link between Floquet-driven topological phase transitions and nonlinear Hall transport.

Off-resonant Floquet engineering provides precisely the perturbation required to address these questions~\cite{zhan2023floquet}. In the high-frequency regime, circularly polarized light generates effective Haldane-like mass terms that modify the Berry curvature and produce Floquet-Bloch bands with tunable topological invariants~\cite{oka2009,kitagawa2011,mikami2016,wangYH2013,claassen2016}. For \(1T'\)-MoS$_2$, recent work has demonstrated that these light-induced mass terms act selectively on different spin-valley sectors, driving the system through a sequence of Floquet topological phase transitions characterized by total Chern numbers \(C_{\mathrm{tot}}=0\rightarrow-1\rightarrow-2\)~\cite{mortezaei2025}. However, it remains unknown how these successive band inversions reconstruct the Berry-curvature dipole and whether the nonlinear Hall response can serve as a direct transport signature of the underlying topological transitions. Although light-induced nonlinear spin Hall currents have recently been investigated in \(1T'\)-WTe$_2$~\cite{bhalla2024light,fei2017} or Floquet graphene systems, neither the sign reversal associated with sequential spin-valley inversions nor the symmetry-imposed selection rules of the \(1T'\) crystal structure have been explored.

In this work, we answer these questions by demonstrating that the intrinsic band geometry of monolayer \(1T'\)-MoS$_2$ gives rise to a unique nonlinear Hall response under Floquet driving. We show that crystal symmetry enforces the selection rule Berry curvature dipole along the $x$ direction \(D_x\equiv0\), while along the $y$ direction a finite \(D_y\) emerges from the intrinsic tilt-induced asymmetry of the electronic bands. More importantly, we demonstrate that each Floquet-induced band inversion reconstructs the momentum-space distribution of the Berry curvature and reverses the sign of both the Berry-curvature dipole and the nonlinear Hall conductivity precisely at the bulk gap-closing points where the Berry curvature reaches its maximum value. In contrast, varying the band tilt alone changes only the magnitude of the nonlinear Hall response without altering its sign, establishing the observed sign reversal as an unambiguous electrical fingerprint of genuine Floquet topological phase transitions. Our results therefore reveal a one-to-one correspondence between nonequilibrium topological transitions and second-order nonlinear transport. More generally, they establish that the \emph{sign}, rather than merely the magnitude, of the nonlinear Hall response provides a universal transport signature of Floquet-induced topological phase transitions. To our knowledge, such an explicit connection between successive Floquet topological transitions and Berry-curvature-dipole sign reversal has not been reported previously.

The remainder of this paper is organized as follows. Section~\ref{sec:model} introduces the low-energy model and derives the effective Floquet Hamiltonian. Section~\ref{sec:transport} presents the Berry curvature, topological phases, Berry-curvature dipole, and nonlinear Hall response as functions of the driving strength, electric field, Fermi energy, and temperature. Finally, Section~\ref{sec:conclusions} summarizes our main findings.

\section{Theory and model}\label{sec:model}

The effective two-band $\mathbf{k}\cdot\mathbf{p}$ Hamiltonian of pristine \(1T'\)-MoS\(_2\) in the presence of a perpendicular electric field can be obtained by unitary transformation to the spin and valley basis~\cite{mortezaei2025,qv3b-ntkc}
\begin{align}
H_{\kappa s}(\bm{k}) =\,&\hbar k_x \nu_1\,\sigma_y
-\hbar k_y\!\left(s\nu_2\,\sigma_x+\kappa\nu_-\,\sigma_0+\kappa\nu_+\,\sigma_z\right)
\nonumber\\
&+\Delta_{\mathrm{so}}(\alpha-s\kappa)\,\sigma_x + V\sigma_0,
\label{eq:Hstatic}
\end{align}
where $\sigma_{x,y,z}$ are the Pauli matrices acting on the orbital pseudospin, $\sigma_0$ is the
identity operator, $s=\pm1$ labels spin, and $\kappa=\pm1$ represents the two valleys. Here
$\Delta_{\mathrm{so}}=41.9$~meV is the intrinsic spin--orbit gap~\cite{mortezaei2025}, $\alpha=|E_z/E_c|$ is the
dimensionless perpendicular field in units of the critical field $E_c=1.42$~V/nm~\cite{qv3b-ntkc,mortezaei2025}, and $V$ is a
gate offset. The anisotropic Dirac velocities and the band tilt are
$(\nu_1,\nu_2,\nu_-,\nu_+)=(3.87,0.46,2.84,7.18)\times10^{5}$~m/s~\cite{mortezaei2025,qv3b-ntkc}. Writing
$H=d_0\sigma_0+\bm{d}\cdot\bm{\sigma}$, the components are:
\begin{align}
d_x &= \Delta_{\mathrm{so}}(\alpha-s\kappa)-\hbar k_y s\nu_2,
& d_y &= \hbar k_x\nu_1, \\
d_z &= -\hbar k_y\kappa\nu_+, & d_0 &= V-\hbar k_y\kappa\nu_-,
\end{align}
and the eigenenergy spectrum and eigenstates follow as
\begin{equation}
\varepsilon^{\pm}_{\kappa s}=d_0\pm|\bm{d}|,\qquad
\psi^{\pm}_{\kappa s}\propto\!\begin{pmatrix} d_x-i d_y\\[2pt]\pm|\bm{d}|-d_z\end{pmatrix},
\label{eq:eps}
\end{equation}
with $|\bm{d}|=(d_x^2+d_y^2+d_z^2)^{1/2}$. In Eq.~\eqref{eq:eps} the spin--orbit term
opens a gap $2\Delta_{\mathrm{so}}|\alpha-s\kappa|$ at $\bm{k}=0$~\cite{mortezaei2025} whose size depends on the spin
and valley indices and on the electric field, while the tilt $\nu_-$ and the gate $V$ enter only
$d_0$ and shift the bands without changing the eigenstates.

\begin{figure*}[t]
\centering
\includegraphics[width=\textwidth]{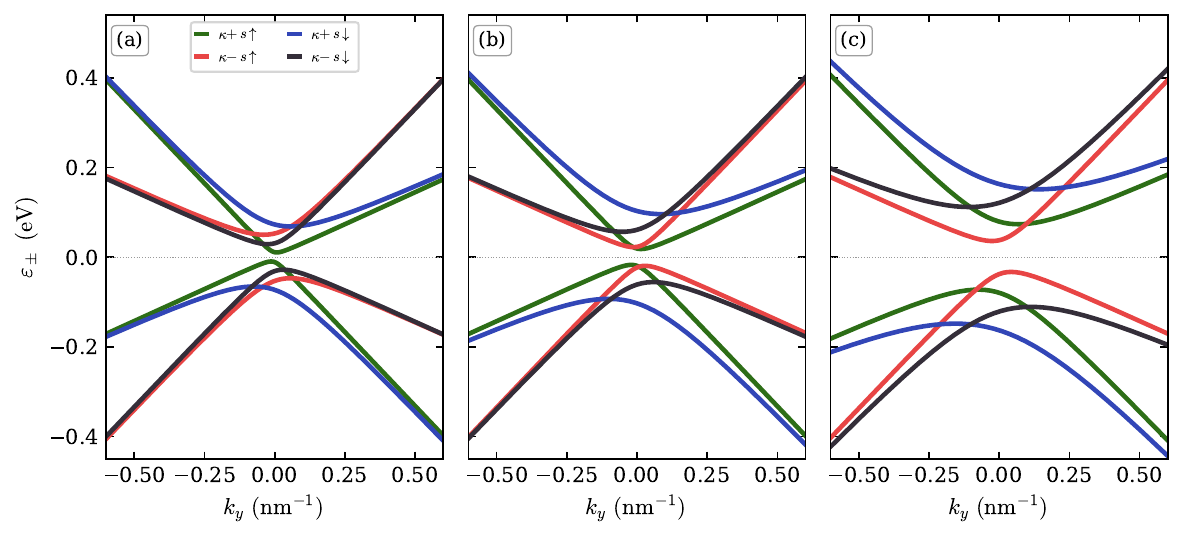}
\caption{\label{fig:bands} Floquet quasienergy bands $\varepsilon_\pm(k_x{=}0,k_y)$ of monolayer
1T$'$-MoS$_2$ for the four spin--valley sectors at three  drive strengths: (a) quantum spin Hall (QSH) (with $C_{\mathrm{tot}}=0$) 
at $\xi=0.01$, (b) spin-polarized quantum Hall insulator (S-QHI) (with $C_{\mathrm{tot}}=-1$) at $\xi=0.04$, and (c) photoinduced quantum Hall insulator (P-QHI) (with $C_{\mathrm{tot}}=-2$) at $\xi=0.10$. Green and blue  are the $\kappa{=}{+}$ spin-up and
spin-down; red and black are  $\kappa{=}{-}$ branches. The electric
field and helicity are fixed at $\alpha=0.5$ and $\eta=+1$.}
\end{figure*}

\subsection{Coupling to a periodic drive}
We consider spatially uniform circularly polarized light in the dipole approximation, described by
the vector potential $\bm{A}(t)=A_0\!\left(\cos\omega t,\ \eta\sin\omega t\right)$ with helicity
$\eta=\pm1$. We implement minimal coupling through $\bm{k}\mapsto\bm{k}+(e/\hbar)\bm{A}(t)$~\cite{vogl2023light}.
Keeping the term linear in $\bm{A}$, the time-dependent Hamiltonian becomes
$H_{\kappa s}(\bm{k},t)=H_{\kappa s}(\bm{k})+e\nu_1 A_x(t)\sigma_y-eA_y(t)\Pi_{\kappa s}$ with
$\Pi_{\kappa s}=s\nu_2\sigma_x+\kappa\nu_-\sigma_0+\kappa\nu_+\sigma_z$. In the off-resonant
regime, where $\hbar\omega$ is larger than the bandwidth and the relevant interband energies, a controlled
high-frequency (van Vleck) expansion yields an effective static Floquet
Hamiltonian~\cite{kitagawa2011,mikami2016,SciPostPhysCore.4.4.033}:
\begin{equation}
H_F = H_{\kappa s}+\frac{1}{\hbar\omega}[H_{-1},H_{+1}]+\mathcal{O}(\omega^{-2}),
\end{equation}
where $H_{\pm1}=\tfrac{eA_0}{2}(\nu_1\sigma_y\mp i\eta\,\Pi_{\kappa s})$ are the $\pm\omega$
Fourier components. Evaluating the commutator with $[\sigma_y,\sigma_x]=-2i\sigma_z$ and
$[\sigma_y,\sigma_z]=2i\sigma_x$ gives
$[H_{-1},H_{+1}]=e^2\nu_1\eta A_0^2(\kappa\nu_+\sigma_x-s\nu_2\sigma_z)$, so that the drive adds~\cite{mortezaei2025}:
\begin{equation}
\delta H_F=\xi\,\eta\,(\kappa\nu_+\sigma_x-s\nu_2\sigma_z),\qquad
\xi\equiv\frac{e^2\nu_1}{\hbar\omega}A_0^2 .
\label{eq:xi}
\end{equation}
The combination $\xi\nu_+$ has the dimension of energy and determines the amplitude of the
light-induced $\sigma_x$ mass. We use this amplitude throughout as the drive control knob and quote it in eV,  denoting it $\xi$ in the figures. In other words, energies are measured in eV,
momenta in nm$^{-1}$, and velocities enter as $\hbar\nu_i$ in eV\,nm. In this normalization the sector inversions discussed below occur at $\xi_{c1}\simeq0.021$ and $\xi_{c2}\simeq0.063$~eV for
$\alpha=0.5$ (and $\xi_{c1}\simeq0.021$, $\xi_{c2}\simeq0.104$~eV for $\alpha=1.5$), in agreement
with Ref.~(\onlinecite{mortezaei2025}). With light-shifted components, the effective Floquet Hamiltonian retains the form
$H_F=d_0\sigma_0+\bm{d}\cdot\bm{\sigma}$ 
\begin{align}
d_x &= M_x-\hbar k_y s\nu_2, & d_y &= \hbar k_x\nu_1, \\
d_z &= M_z-\hbar k_y\kappa\nu_+, & d_0 &= V-\hbar k_y\kappa\nu_- ,
\label{eq:d}
\end{align}
and two drive-induced effective masses,
\begin{equation}
M_x=\Delta_{\mathrm{so}}(\alpha-s\kappa)+\xi\nu_+\eta\kappa,\qquad
M_z=-\xi\nu_2\eta s .
\label{eq:masses}
\end{equation}

As will be shown below, the dependence of $M_x$ and $M_z$ on $\xi$, $s$, and $\kappa$ plays a crucial role in determining the physical properties discussed in this work.

Furthermore, the light contributes a Haldane-like $\sigma_x$ mass that renormalizes the inversion gap and a spin-selective $\sigma_z$ mass~\cite{hasan2010colloquium}. We find the light-shifted critical field at which a given spin-valley sector inverts by setting $M_x=0$ at
$\alpha_c^{(\kappa s)}=s\kappa-(\xi\nu_+/\Delta_{\mathrm{so}})\kappa\eta$ ~\cite{mortezaei2025}.
\begin{figure*}[htbp]
\centering
\includegraphics[width=\textwidth]{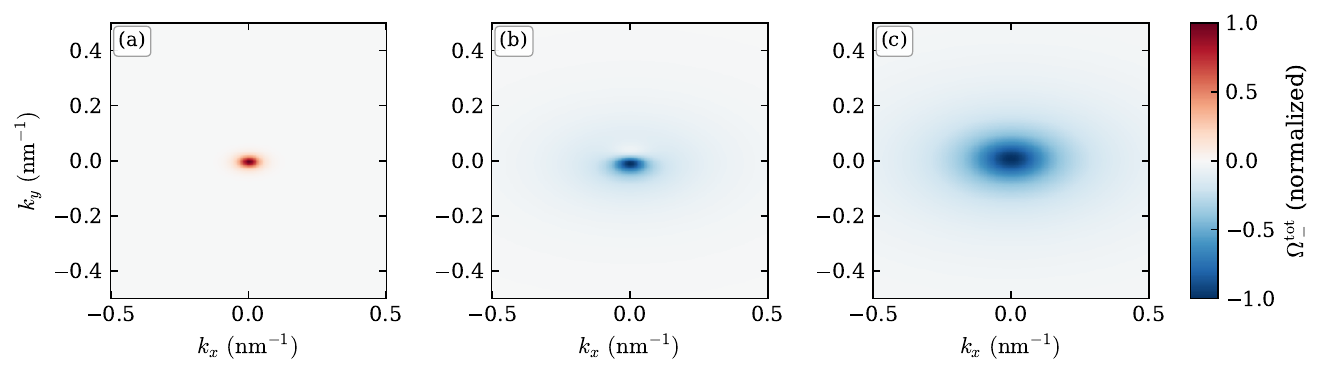}
\caption{\label{fig:berrytot} Total occupied-band Berry curvature
$\Omega^{\rm tot}_-=\sum_{\kappa,s}\Omega^z_-$ of monolayer 1T$'$-MoS$_2$ versus $k_x$ and $k_y$,
for (a) QSH ($\xi=0.01$), (b) S-QHI ($\xi=0.04$), and (c) P-QHI ($\xi=0.10$). Each panel is normalized to its own peak
value to make the sign visible. Other parameters are as in Fig~(\ref{fig:bands}). The Berry curvature is noticeably peaked at the center of the band, due to the minimum value of $|{\bf{d}}|$ and its shape
is determined by gap value. }
\end{figure*}

\begin{figure*}[htbp]
\centering\includegraphics[width=\textwidth]{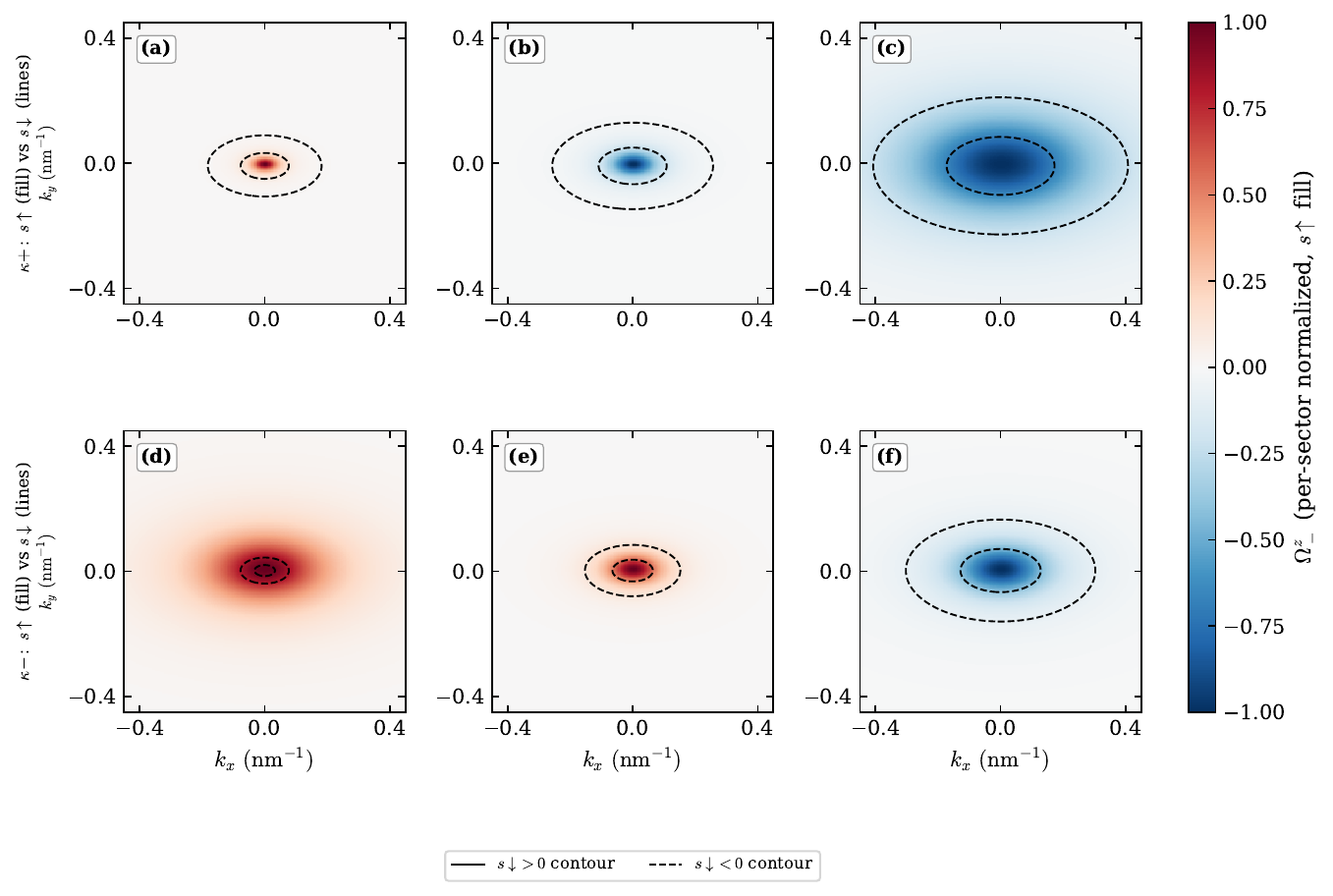}
\caption{\label{fig:berrysec} Sector-resolved occupied-band Berry curvature $\Omega^z_-$. Each row is one valley (top) $\kappa{+}$, (bottom) $\kappa{-}$ and each column one drive strength (a,d) QSH ($\xi=0.01$), (b,e) S-QHI ($\xi=0.04$), (c,f) P-QHI ($\xi=0.10$). Filled color shows $s{\uparrow}$; black contours overlay $s{\downarrow}$ (solid: positive, dashed: negative), each normalized to its own peak. Other parameters are as in Fig.~(\ref{fig:bands}).}
\end{figure*}

\smallskip

In Fig.~(\ref{fig:bands}), we show how circularly polarized light reshapes the
band structure of monolayer \(1T'\)-MoS\(_2\) through values of $M_x$ and $M_z$. Panels \subred{fig:bands}{a-c} show the Floquet quasienergy bands of the four
spin--valley sectors at three drive strengths, \(\xi=0.01\), \(0.04\), and
\(0.10\). The red and blue bands are the spin-up and spin-down branches of
the \(\kappa=+\) valley, and the orange and black bands are the same for the
\(\kappa=-\) valley. Even before the light is turned on, the bands are tilted
about \(k_y=0\), because the \(\kappa\nu_-\) term shifts the two bands of each
sector together without changing the gap. This tilt is important later because
it is what gives \(1T'\)-MoS\(_2\) an intrinsic Berry-curvature dipole. In the case of a weak drive Fig.~\subred{fig:bands}{a}, the four bands are almost overlapping and only the small spin–orbit gap is visible, so the system looks like the ordinary quantum spin Hall insulator (QSH). As the drive strength increased in Fig.~\subred{fig:bands}{b-c}, the
light-induced masses act differently on each spin and valley, so the gap of
one channel at a time first shrinks almost to zero and then reopens with the band order flipped. The key outcome of this figure is that the light does not
shift all the bands together. It opens and closes gaps selectively, one spin
and valley channel at a time, and each of these closings is the starting point
of a topological transition that we later detect directly in the nonlinear
Hall response.

\section{Transport characteristics}\label{sec:transport}
\subsection{Berry curvature and topological phases}
For a two-band Hamiltonian $H=d_0\sigma_0+\bm{d}\cdot\bm{\sigma}$ the Berry curvature of the
upper ($+$) and lower ($-$) bands is
$\Omega^z_\lambda=-\tfrac{\lambda}{2}\,\bm{d}\cdot(\partial_{k_x}\bm{d}\times\partial_{k_y}\bm{d})/
|\bm{d}|^3$; the scalar $d_0$ does not contribute. With
$\partial_{k_x}\bm{d}=(0,\hbar\nu_1,0)$ and $\partial_{k_y}\bm{d}=(-\hbar s\nu_2,0,-\hbar\kappa\nu_+)$
the numerator is independent of $\bm{k}$, and the curvature of the occupied (lower) band reduces to~\cite{mortezaei2025}
\begin{equation}
\Omega^z_-(\bm{k})=-\frac{\hbar^2\nu_1}{2}\,
\frac{\kappa\nu_+M_x-s\nu_2M_z}{|\bm{d}|^3},\qquad \Omega^z_+=-\Omega^z_- .
\label{eq:berry}
\end{equation}
Equation~\eqref{eq:berry} shows that the curvature is localized around $\bm{k}=0$, where $|{\bf{d}}|$ gets a minimum, and its sign is determined only by the light-induced masses $M_x$ and $M_z$, while the tilt $\nu_-$ and the potential $V$ do not play any role. The sector Chern number is obtained by integrating the occupied-band curvature over the plane gives
\begin{equation}
C_{\kappa s}=\frac{1}{2\pi}\!\int\! d^2k\,\Omega^z_-
=-\tfrac12\,\mathrm{sgn}\!\left(\kappa\nu_+M_x-s\nu_2M_z\right),
\label{eq:chern}
\end{equation}
 the half-integer Chern number of the filled band that contributes to the Hall response, summing over sectors, gives the total Chern number $C_{\mathrm{tot}}=\sum_{\kappa,s}C_{\kappa s}$, along with
the spin- and valley-resolved partial sums $C_s=\sum_\kappa C_{\kappa s}$ and
$C_\kappa=\sum_s C_{\kappa s}$\cite{mortezaei2025}.

\smallskip
The key physical implication is the total Berry curvature of the filled
bands summed over all four spin-valley sectors shown in Fig.~(\ref{fig:berrytot}). This is the first place where
the light's effect on the topology is clearly observed. Each panel in
Fig.~(\ref{fig:berrytot}) is normalized to its own peak, so the sign remains
readable even as the overall magnitude varies from one drive strength to the
next. The curvature is sharply peaked at the center of the band, where $|{\bf{d}}|$ gets a minimum, and its shape
is almost entirely determined by whichever sector has the smallest gap at that
drive. A small gap gives rise to a tall, narrow peak, whereas the strongly
gapped sectors contribute only a low and broad background hardly visible
against it. The colour of this dominant peak identifies the physics. In the
quantum spin Hall phase \subred{fig:berrytot}{a}, it is positive and shown red, but once the
first channel inverts \subred{fig:berrytot}{b}, it flips to negative and turns blue, and it
stays negative as the drive is increased further \subred{fig:berrytot}{c}. The important
subtlety is that this local colour only follows the smallest-gap channel. The
real topological information is not in the peak at all. The domain is the whole
area under the curvature which steps through \(C_{\rm tot}=0,-1,-2\) when the
sectors are inverted one after the other. Viewed like this, the light does
more than rescaling the Berry curvature. It reconstructs the curvature, sector
by sector, and this controlled reconstruction then gives rise to a tunable
sign-changing nonlinear Hall response.

The origin of the sign of the nonlinear Hall response is shown in
Fig.~(\ref{fig:berrysec}). The black contours give the \(s\downarrow\), at a
fixed valley, across the three drive strengths. The filled colour gives the
\(s\uparrow\) curvature. In the \(\kappa+\) valley, the \(s\uparrow\) curvature
(top row) changes from positive (red) to negative (blue) between
\subred{fig:berrysec}{a-b}, at \(\xi_{c1}\). The same flip occurs later
in the bottom row of the \(\kappa-\) valley, between \subred{fig:berrysec}{e-f}
, at \(\xi_{c2}\). The \(s\downarrow\) contours remain negative
throughout. Each flip is a band inversion that changes the Chern number of that
sector by one unit. More importantly for what follows, each flip changes the
sign of that sector's Berry curvature, and it is that reversal which the
Berry-curvature dipole later inherits, so the curvature flips here are the
direct origin of the sign changes in the nonlinear Hall response, and the $M_x$ values control it.

%\begin{figure*}[htbp]
%\centering
%\includegraphics[width=\textwidth]{YF3a_berry_kx.pdf}
%\caption{\label{fig:berrycutx} Line cuts of the occupied-band Berry curvature $\Omega^z_-$ of the
%four spin--valley sectors versus $k_x$ at fixed $k_y=0.05~\mathrm{nm}^{-1}$, (a) QSH ($\xi=0.01$), (b) S-QHI ($\xi=0.04$), (c) P-QHI ($\xi=0.10$) Other parameters are as in Fig.~(\ref{fig:bands}).}
%\end{figure*}

%\begin{figure*}[htbp]
%\centering
%\includegraphics[width=\textwidth]{YF3b_berry_ky.pdf}
%\caption{\label{fig:berrycuty} Same as Fig.~(\ref{fig:berrycutx}) but versus $k_y$ at fixed
%$k_x=0.10~\mathrm{nm}^{-1}$. The curvature is asymmetric about $k_y=0$, in contrast to the
%symmetric $k_x$ cuts.}
%\end{figure*}

Only selected channels flip because of the light-shifted mass
\(M_x = \Delta_{\rm so}(\alpha - s\kappa) + \xi\nu_+\eta\kappa\). For helicity
\(\eta = +1\) the drive can cancel the equilibrium mass only in the two spin-up
sectors, so only their curvature is free to change sign. There is a drive in
the spin-down sectors, which increases their mass instead of opposing it, so
they never flip and contribute a fixed background. The two valleys also respond
to different drives, just because of the gap they start from. For
\(\alpha < 1\), the \(\kappa^+\) valley has the smaller gap, and is therefore
flipped first by the light at \(\xi_{c1}\) and only later the \(\kappa^-\)
valley at \(\xi_{c2}\). This ordered reversal channel-by-channel is important
for the nonlinear response. 
Each inverted sector produces a Berry curvature of opposite sign, and the nonlinear
Hall signal switches sign along with the transitions
\(C_{\rm tot} = 0 \to -1 \to -2\).
The figure thus establishes a direct connection between the microscopic band inversions and the sign-changing nonlinear Hall response.

\subsection{Berry-curvature dipole}
The nonlinear Hall response is governed by the dipole moment of the Berry curvature, defined for
band $\lambda$ as~\cite{yar2022,PhysRevLett.121.246403}
\begin{equation}
D_i=\int\!\frac{d^2k}{(2\pi)^2}\,(\partial_{k_i}\Omega_\lambda)\,f_\lambda
=-\int\!\frac{d^2k}{(2\pi)^2}\,\Omega_\lambda\,\frac{\partial f}{\partial\varepsilon}\,
\partial_{k_i}\varepsilon_\lambda ,
\label{eq:bcd}
\end{equation}
where $f$ is the Fermi--Dirac distribution and $i$ represents $x$ or $y$ coordinate. The total dipole is a sum of contributions from both Floquet bands, $D_i = \sum_{\lambda=\pm} D_i^{\lambda}$
; whichever band the Fermi level crosses responds, and $D_y$ remains finite on both sides of the gap. The curvature is even in $k_x$ and $\partial_{k_x}\Omega$ is odd, since $k_x$ enters the spectrum and the curvature only via
 $k_x^2$, so the $x$
component of the dipole vanishes identically, $D_x\equiv0$. The linear-in-$k_y$ terms in $d_x$ and
$d_z$ make $\Omega$ asymmetric about $k_y=0$, so $\partial_{k_y}\Omega$ has a nonzero integral and
$D_y\neq0$.
This finite Berry curvature dipole $D_y$ originates from the anisotropic, tilted bands of the 1T$'$ structure; hence, there is a non-linear Hall response for the tilted Dirac fermions of \(1T'\)-MoS\(_2\) without any hexagonal warping of the Fermi surface.

\begin{figure*}[htbp]
\centering
\includegraphics[width=\textwidth]{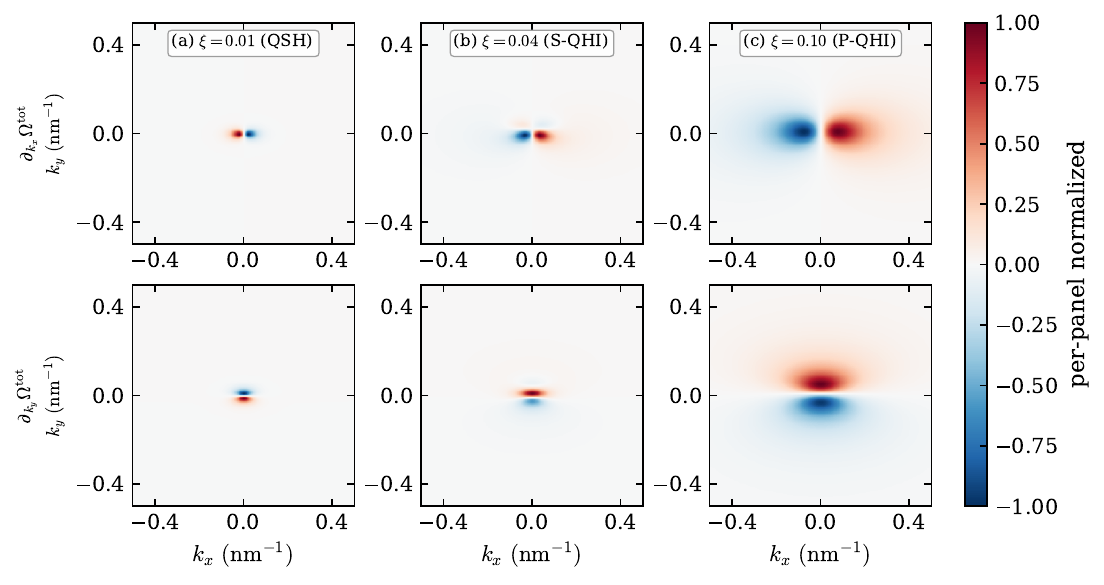}
\caption{\label{fig:bcdmap} Berry-curvature-dipole density of monolayer \(1T'\)-MoS\(_2\) summed over
sectors: (top) $\partial_{k_x}\Omega^{\rm tot}$ and (bottom) $\partial_{k_y}\Omega^{\rm tot}$, for
(a) $\xi=0.01$, (b) $\xi=0.04$, and (c) $\xi=0.10$, each panel normalized to its own peak.
$\partial_{k_x}\Omega$ is odd in $k_x$ and integrates to zero ($D_x=0$), while
$\partial_{k_y}\Omega$ integrates to a finite $D_y$. Other parameters are as in
Fig.~(\ref{fig:bands}).}
\end{figure*}

The physical feature of the BCD density summed over all sectors, each
normalized to its own peak, is shown in Fig.~(\ref{fig:bcdmap}). In the top row, \(\partial_{k_x}\Omega\) is an odd
function of \(k_x\), with equal and opposite lobes that integrate to zero,
while in the bottom row, \(\partial_{k_y}\Omega\) is asymmetric about
\(k_y = 0\) and integrates to a finite value. Both densities are concentrated
near the band center. So only \(\partial_{k_y}\Omega\) can create a finite
dipole, while \(\partial_{k_x}\Omega\) enforces \(D_x \equiv 0\).

These parities are inherited directly from the symmetry of the curvature
itself. Since \(\Omega^z_-\) is an even function of \(k_x\) (\(k_x\) enters
only through the \(k_x^2\) term of \(|\bm{d}|\)), its derivative
\(\partial_{k_x}\Omega\) is necessarily odd in \(k_x\), which explains why the
top-row maps show two lobes of equal magnitude and opposite sign that cancel
exactly upon integration. On the other hand, along the \(k_y\) the curvature is
asymmetric so that \(\partial_{k_y}\Omega\) does not separate into cancelling
lobes and keeps a net weight. This figure thus illustrates in momentum space
the origin of the selection rule \(D_x \equiv 0\), \(D_y \neq 0\): the dipole,
as the first moment of the curvature, can be nonzero only in the direction in
which the curvature does not have a definite parity. For \(1T'\)-MoS\(_2\),
this direction is \(k_y\), determined by the band tilt and the
\(\kappa\nu_+\) term.

The figure also reveals that the two density components are sharply localized
around \(\bm{k} = 0\), in the same region where the Berry curvature is peaked,
and that their amplitude increases from the QSH panel to the P-QHI panel as
the drive reshapes the curvature. Even where \(\partial_{k_x}\Omega\) has large
local values, its odd structure guarantees a vanishing integral, so a strong
density does not mean a strong dipole. For the nonlinear Hall response, it is
the asymmetric part, given by \(\partial_{k_y}\Omega\), that is important. The
concentration of \(\partial_{k_y}\Omega\) around the band centre provides an
additional explanation for the fact that the dipole, and thus the response, is
largest when the Fermi level is close to the gap. This is reflected in the
energy dependence studied in the following figures.
We would like to highlight that

\begin{eqnarray}
\partial_{k_y}\Omega^z_-(\mathbf{k})
=&&
-\frac{3\hbar^3 \nu_1}{2}\,
\frac{
(\kappa \nu_+ M_x - s \nu_2 M_z)
}{
|\mathbf{d}|^5
}\nonumber\\
&&\times\Big[
s\nu_2 (M_x - \hbar k_y s\nu_2)
+
\kappa \nu_+ (M_z - \hbar k_y \kappa \nu_+)
\Big],\nonumber\\
\end{eqnarray}
can be understood as a measure of the momentum-space asymmetry of a Floquet-renormalized Berry curvature field. Because the Berry curvature has the form $\Omega_-^z \propto |\mathbf{d}|^{-3}$, where $\mathbf{d}$ functions as an effective Bloch vector, its $k_y$-derivative does not change the monopole strength but instead examines how spin-orbit and exchange-induced terms cause the center of this geometric structure to shift in momentum space. Specifically, the reliance of $d_x$ and $d_z$ on $k_y$ suggests that changing $k_y$ translates the effective field $\mathbf{d}$ in a nonuniform way, resulting in a geometric response controlled by the projection $\mathbf{d}\cdot\partial_{k_y}\mathbf{d}$. Consequently, $\partial_{k_y}\Omega_-^z$ gains a distinctive $|\mathbf{d}|^{-5}$ enhancement, suggesting that the response is dramatically peaked close to avoided band crossings where the effective gap is small. This variable directly measures the first moment of a shifted Berry curvature monopole and physically encodes the redistribution of Berry curvature in momentum space under anisotropic Floquet-induced band distortions. The underlying symmetry-breaking chirality of the Floquet-dressed system is reflected in the prefactor $(\kappa\nu_+ M_x - s\nu_2 M_z)$. The remaining structure controls the extent to which this asymmetry is heightened near regions of strong band hybridization.

The companion component \(\partial_{k_x}\Omega\), on the other hand, is shown
to retain its antisymmetric form in every phase, so that its integral over the
Brillouin zone vanishes identically, and the drive can never generate an
\(x\)-directed dipole; the exact result \(D_x \equiv 0\) thus holds in all
three phases, not just at equilibrium.

%This figure, together with Figs.~(\ref{fig:bcdmap}) and (\ref{fig:bcdcutx}),
%completes the momentum-space picture of the dipole: the nonlinear Hall
%response of \(1T'\)-MoS\(_2\) is entirely due to the \(k_y\)-asymmetric part of
%the Berry curvature from the tilted, anisotropic bands, while the
%\(k_x\)-even part is prohibited from contributing by symmetry.

\subsection{Relaxation time and nonlinear Hall response}
The second-order Hall response within a constant relaxation-time approximation and low-frequency limit is~\cite{yar2022,ortix2021}
\begin{equation}
\chi_{lmn}=\frac{e^3\tau}{\hbar^2}\!\int\!\frac{d^2k}{(2\pi)^2}\,
\epsilon_{lmp}\,(\partial_{k_n}\Omega_p)\,f_{\bm{k}}
\;\Longrightarrow\;
\chi_{xyy}=\frac{e^3\tau}{\hbar^2}\,D_y ,
\label{eq:chi}
\end{equation}
The response is directly proportional to the Berry-curvature dipole. The relaxation time
is obtained, for $\delta$-correlated spin-independent disorder
$V_{\rm imp}=\sum_i V_i\delta(\bm{r}-\bm{R}_i)$ with $\langle V_i^2\rangle=V_0^2$ and impurity
density $n_i$, from~\cite{yar2022}
\begin{equation}
\frac{1}{\tau}=\frac{2\pi}{\hbar}n_iV_0^2\!\int\!\frac{d^2k'}{(2\pi)^2}
(1-\cos\theta_{\bm{k}\bm{k}'})\,\delta(E_F-\varepsilon_{\bm{k}'}) ;
\end{equation}
in what follows, for the sake of simplicity, we take a constant $\tau=0.1$~ps, which only rescales the overall magnitude. In numerical results, we will only plot $D_y$ because $\chi_{xyy}$ and $\sigma_{xy}$ are defined by $D_y$.

The
nonlinear Hall conductivity in an applied field $\bm{E}$ is $\sigma_{xy}=2\chi_{xyy}E$~\cite{ma2019observation,du2019disorder}.

\begin{figure*}[htbp]\label{DEF}
\centering
\includegraphics[width=\textwidth]{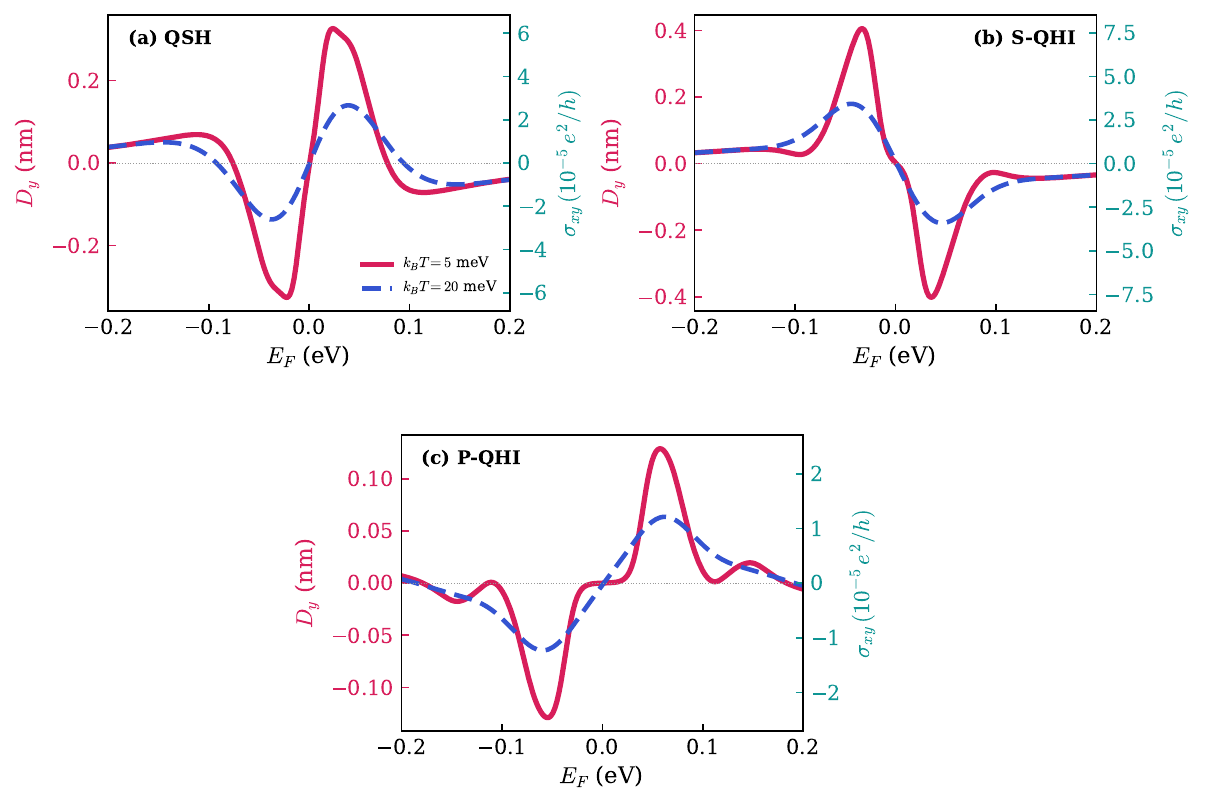}
\caption{Berry-curvature dipole \(D_y\) (left axis) and nonlinear Hall
conductivity \(\sigma_{xy}=2\chi_{xyy}E\) (right axis) of monolayer
\(1T'\)-MoS\(_2\) versus the Fermi energy, for (a)~QSH (\(\xi=0.01\)),
(b)~S-QHI (\(\xi=0.04\)), and (c)~P-QHI (\(\xi=0.10\)). Pink curves are for 
\(k_BT=5\)~meV, dashed blue curves are for \(k_BT=20\)~meV. \(D_x\equiv 0\) by \(k_x\)
parity and is not shown. Since \(\chi_{xyy}=(e^3\tau/\hbar^2)D_y\), the two
quantities share the same curve.  \(\tau=0.1\)~ps and \(E=100\)~V/m. Other
parameters are as in Fig.~(\ref{fig:bands}).}
\label{fig:DEf}
\end{figure*}

The Berry-curvature dipole \(D_y\) together with
the nonlinear Hall conductivity \(\sigma_{xy}\) as a function of the Fermi
energy , integrated over the full Brillouin zone, at two different temperatures are shown in Fig.~(\ref{fig:DEf}). Since
\(\chi_{xyy}=(e^3\tau/\hbar^2)D_y\) and \(\sigma_{xy}=2\chi_{xyy}E\) differ from
\(D_y\) only by constant prefactors, the three quantities share the same
lineshape and sign structure. We therefore take \(D_y\) as the primary quantity
(left axis) and display the measurable \(\sigma_{xy}\) on the right axis, and we
do not illustrate \(\chi_{xyy}\) separately. In addition to the sector
inversions, the sign of the central feature itself switches between the QSH,
S-QHI and P-QHI phases, tracking the sector inversions. Increasing \(k_B T\) broadens the Fermi-Dirac distribution, leading to a monotonic suppression of the nonlinear Hall response as the thermally broadened occupation averages the dispersive Berry curvature over a wider energy window.

The dispersive lineshape follows naturally from \(D_y\) being a Fermi-surface
quantity. The dipole is given by
\(D_y=-\int \frac{d^2k}{(2\pi)^2}\,\Omega\,(\partial f/\partial\varepsilon)\,
\partial_{k_y}\varepsilon\), where the factor \(\partial f/\partial\varepsilon\)
restricts the integral to a thin shell at the Fermi level, so that \(D_y(E_F)\)
probes the product of the Berry curvature and the band velocity right where the
Fermi surface sits. When \(E_F\) is slightly below the gap, the shell wraps the
top of the valence band, and when it is slightly above the gap, it wraps the
bottom of the conduction band. The two bands carry opposite curvature, so
\(D_y\) takes opposite signs on the two sides and passes through zero inside the
gap, where there is no Fermi surface. Because the curvature is sharply peaked at
the band center, the response is largest when \(E_F\) touches the gap edge and
falls off as the Fermi level is driven deep into either band. The summation over
both Floquet bands is essential here. It is what makes \(D_y\) finite on both
sides of the gap rather than only on the side that happens to be occupied.
Because \(\sigma_{xy}=2\chi_{xyy}E\) is of second order in the applied field, it
depends on \(E\) and vanishes in the linear limit, in contrast to the ordinary
Hall conductivity. For fixed drive, \(\sigma_{xy}\) scales linearly with \(E\),
and since \(D_x=0\) the response is purely transverse. Its sign change across the gap and
between the phases is exactly the second-harmonic Hall signal that a transport
experiment would record, so that \(\chi_{xyy}\) and equivalently
\(\sigma_{xy}\) provides an experimentally accessible label of the phase.

We find the BCD ($D_y$) is of order 3\text{--}4\AA, comparable to the intrinsic dipole estimated for monolayer and bilayer WTe$_2$, which is of order 10\AA ~\cite{PhysRevLett.121.266601,ma2019} but it is smaller than the value reported for the engineered moir\'e system, such as twisted bilayer WTe$_2$ (1400\AA)~\cite{He_2021_Giant_NHE_WTe2}. Our result is therefore comparable to the  WTe$_2$, rather than to the engineered giants, but with the distinguishing feature that here the dipole and hence the sign of the nonlinear Hall response are generated and reversed optically by the circularly polarized drive.

\begin{figure}[htbp]
\centering 
\includegraphics[width= 1\columnwidth]{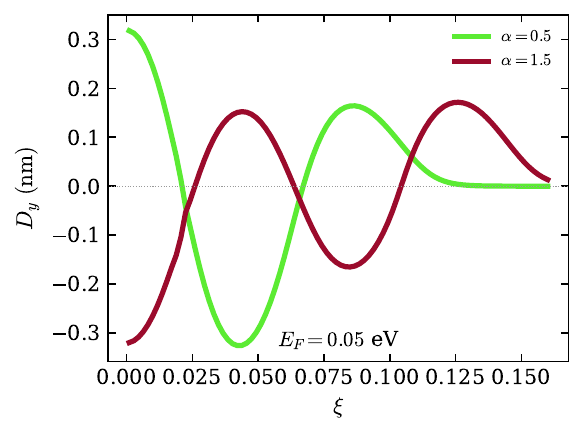} 
\caption{\label{fig:bcdxi} Berry-curvature dipole $D_y$ versus the Floquet coupling strength $\xi$ at two electric fields, $\alpha=0.5$ (blue) and $\alpha=1.5$ (red), for $E_F=0.05$~eV and $k_BT=5$~meV.} 
\end{figure}

\begin{figure}[htbp]
\centering
\includegraphics[width=1\columnwidth]{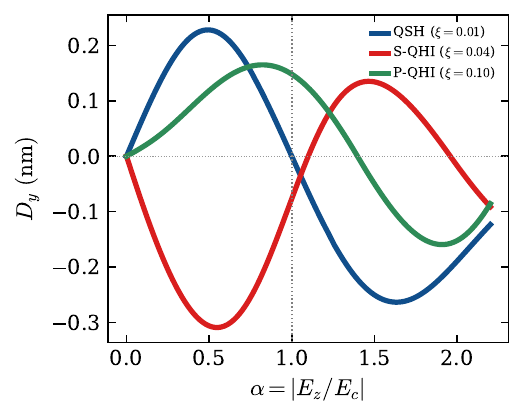}
\caption{\label{fig:vsalpha} Berry-curvature
dipole $D_y$ versus the electric field $\alpha=|E_z/E_c|$ for the three driven phases QSH
($\xi=0.01$), S-QHI ($\xi=0.04$), P-QHI ($\xi=0.10$). The dotted vertical line marks $\alpha=1$.
Here $E_F=0.05$~eV and $k_BT=5$~meV.}
\end{figure}

\begin{figure*}[htbp]
\centering\includegraphics[width=1\textwidth]{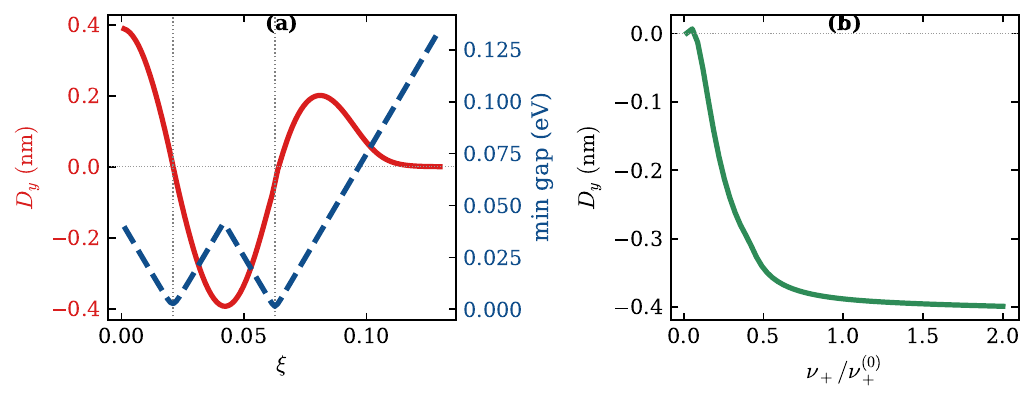}
\caption{\label{fig:osc} (a) Berry-curvature dipole $D_y$ (red, left axis) and minimum
quasienergy gap (blue dashed, right axis) versus the Floquet coupling strength $\xi$; (b)
$D_y$ versus the band-tilt strength $\nu_+/\nu_+^{(0)}$. Here $E_F=0.04$~eV, $\alpha=0.5$, and $k_BT=5$~meV.}
\end{figure*}

Figure~(\ref{fig:bcdxi}) shows the Berry-curvature dipole \(D_y\) as a function
of \(\xi\) at the two electric fields. This figure shows that the sharp features of
\(D_y\) occurs at critical drives \(\xi \simeq 0.021\) and \(0.063\), confirming that the sign structure of \(\chi_{xyy}(\xi)\)
is inherited directly from the dipole, and that the upper transition moves to
larger \(\xi\) as the static field \(\alpha\) is increased. Because \(D_y\)
contains none of the prefactors of Eq.~\eqref{eq:chi}, neither the relaxation
time \(\tau\) nor the carrier charge, this figure isolates the purely geometric
part of the response and shows that both the bipolar features and their sign
are properties of the driven band structure alone, not artifacts of the
scattering model. The coincidence of the feature positions in
%Figs.~(\ref{fig:chixi}) and (\ref{fig:bcdxi}) therefore establishes that the
~(\ref{fig:bcdxi}) therefore establishes that the
constant-\(\tau\) approximation only rescales the magnitude of $D_y$ and hence  \(\chi_{xyy}\)
and leaves the topology-driven sign reversals untouched, so that the location
of every transition can be read off equally well from the dipole or from the measurable response. This is what makes the sign reversal a robust signature:
because it originates in the band geometry rather than in disorder or material
details, the nonlinear Hall response marks the Floquet topological transitions
independently of the scattering that sets its overall scale.

In theory, the nonlinear Hall conductivity does not measure the integral of the Berry-curvature distribution, but rather its initial moment. By using band inversion to recreate this distribution, a Floquet topological transition reverses the dipole moment's orientation. As a result, while the fundamental symmetry of the lattice does not change, the nonlinear Hall signal does.

\begin{figure*}[htbp]
\centering\includegraphics[width=\textwidth]{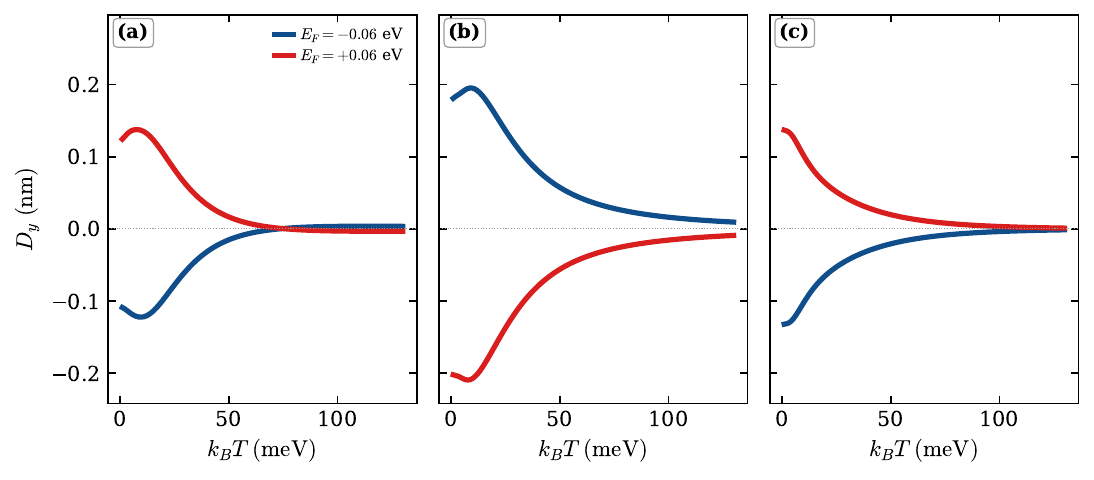}
\caption{\label{fig:chiT} Berry-curvature dipole $D_y$  versus temperature for
$E_F=-0.06$~eV (below the gap, blue) and $E_F=+0.06$~eV (above the gap, red), for (a) QSH
($\xi=0.01$), (b) S-QHI ($\xi=0.04$), and (c) P-QHI ($\xi=0.10$). Other parameters are as in Fig~(\ref{fig:bands}).}
\end{figure*}

The Berry-curvature dipole \(D_y\) as a
function of the electric field \(\alpha=|E_z/E_c|\) for the three driven phases is shown in Fig.~(\ref{fig:vsalpha}).
Our results show that each phase reverses the sign of its response at its own
light-shifted critical field
\(\alpha_c^{(\kappa s)}=s\kappa-(\xi\nu_+/\Delta_{\mathrm{so}})\kappa\eta\), so
that the curves of the different phases cross zero at different values of
\(\alpha\) on either side of \(\alpha=1\). The electric field and the drive
therefore act together to determine the susceptibility of each spin--valley
sector to the mass renormalization. The field-tuned sign reversals originate in
the Berry-curvature dipole itself rather than in the relaxation time or the
applied field. This establishes the perpendicular electric field as a second,
all-electrical knob that controls the sign of the nonlinear Hall response,
complementary to the optical drive and acting through the same underlying
dipole.

The physical link between the sign changes and the
topological transitions is illustrated explicitly in Fig.~(\ref{fig:osc}). The zero crossings and
sharp features of $D_y$(\(\xi)\) coincide with the minima of the
quasienergy gap (blue dashed) at \(\xi_{c1}\) and \(\xi_{c2}\) is shown in panel (a), so that every
sign reversal of the nonlinear Hall response is accompanied by a bulk gap
closing. Results in Panel~(b), by contrast, sweep the band tilt \(\nu_+\), which sets the
\(k_y\) asymmetry of the curvature-carrying part of the Hamiltonian and thereby
generates and tunes the Berry-curvature dipole, but induces no band inversion.
The magnitude of $D_y$ grows monotonically with \(\nu_+\) and then saturates, and it never
changes sign. The dipole is already finite at small tilt, because it originates
in the band anisotropy, and increasing \(\nu_+\) simply amplifies it until the
curvature redistribution saturates. The contrast between the two panels is the
key point: a sign reversal is a specific signature of a Floquet-induced
topological transition, whereas a parameter that reshapes the bands without
inverting them only rescales the response. This is what lets the sign of the
nonlinear Hall signal, not merely its magnitude, serve as an unambiguous marker
of the drive-induced transitions.

Finally, the berry-curvature dipole as a function of
temperature for Fermi energies below and above the gap is illustrated in Fig.~(\ref{fig:chiT}). At \(T \to 0\) the
dipole is already finite, because \(E_F = \mp 0.06\)~eV lies inside the bands
rather than in the gap, so a Fermi surface, and hence a Berry-curvature dipole,
exists even without thermal excitation. In the QSH and
S-QHI phases the magnitude of the dipole first
grows slightly with temperature, as the broadening thermal window draws in
additional states near the gap edge where the Berry curvature is large, reach
an extremum at low-to-moderate \(k_BT\), and then decays as further broadening
averages the dispersive curvature toward zero as shown in Fig.~\subred{fig:chiT}{a-b}. In the P-QHI phase
Fig.~\subred{fig:chiT}{c}, the gap is widest due to the $M_z$ values, and the curvature least
concentrated, so the response is already maximal at \(T \to 0\) and decays monotonically with temperature by changing the Fermi-Dirac distribution, without an intermediate maximum.

Because \(D_y(E_F)\) is dispersive and approximately odd about the gap center,
the below-gap and above-gap curves are near mirror images,
\(\chi_{xyy}(-E_F) \simeq -\chi_{xyy}(E_F)\), so that the two curves in each
panel straddle zero and relax toward small values at high temperature. The
decay at high \(k_BT\) is common to all three phases and reflects the same
mechanism: a thermal window broad compared with the gap samples the positive
and negative lobes of the dispersive dipole together, so their contributions
increasingly cancel. The temperature dependence therefore, arises from the
competition between the thermal population of the high-curvature band-edge states,
which enhances the response, and thermal smearing of the dispersive dipole,
which suppresses the dipole. The competition leads to an intermediate-temperature maximum
in the more weakly gapped phases in Fig.~\subred{fig:chiT}{a-b} and a monotonic falloff in the
strongly gapped phase in Fig.~\subred{fig:chiT}{c}.  Practically, this places the largest nonlinear
Hall signal at low temperature with the Fermi level tuned close to a band edge,
which is the regime most favorable for detecting the drive-controlled response
in experiment.

We should emphasis that increasing the temperature suppresses the nonlinear Hall response, making the effect strongest at low temperatures and progressively weaker as thermal broadening increases. Moreover, the response also survives in the presence of
impurities. For the
tilted Dirac model relevant to \(1T'\)-MoS\(_2\), a full quantum treatment
showed \cite{Du_2021} that disorder contributions enhance the magnitude of the nonlinear
Hall conductivity. Moderate disorder is anticipated to widen but not completely eradicate the transition since the predicted sign reversal is linked to a bulk topological transition rather than a quantitative amplification of the Berry-curvature dipole, so long as the mobility gap stays open. Thus, the sign reversal is robust against disorder and finite temperature.

Before summarizing the results, we note that current nonlinear Hall measurement techniques enable our predictions to be directly tested. Berry-curvature-dipole transport in 2D materials has already been electrically detected by second-harmonic Hall investigations. Under off-resonant circularly polarized illumination, the carrier density and perpendicular electric field are independently controlled by electrostatic gates, while the laser intensity offers a continuous tuning parameter that drives successive Floquet topological transitions. The expected reversal of the second-harmonic Hall voltage at the crucial optical intensities corresponding to Floquet gap closings is the most characteristic experimental signature. The sign change offers a reliable transport fingerprint of nonequilibrium topology since it is associated with band inversion rather than the Berry-curvature dipole's amplitude.
\smallskip

\section{Conclusions}\label{sec:conclusions}

We have shown that off-resonant circularly polarized light is an efficient and tunable route 
to optically control the nonlinear Hall response of monolayer \(1T'\)-MoS\(_2\) through 
Floquet-driven topological phase transitions. Our calculations show that the symmetry selection 
rule absent in trigonal TMDs is enforced by the anisotropic tilted band structure of the 
\(1T'\) lattice: the $x$-component of the Berry-curvature dipole vanishes identically 
($D_x \equiv 0$), while a finite $D_y$ arises solely from the $k_y$-asymmetry of the tilted 
bands, rather than from trigonal warping or strain. The off-resonant drive induces spin- and 
valley-selective Haldane-like mass corrections, sequentially inverting individual spin-valley 
sectors and driving the system through quantum spin Hall ($C_\mathrm{tot} = 0$), spin-polarized 
quantum Hall ($C_\mathrm{tot} = -1$), and photoinduced quantum Hall ($C_\mathrm{tot} = -2$) 
phases.

The sector-by-sector reconstruction of the Berry curvature at each band inversion and the exact 
sign reversal of $D_y$ and the nonlinear Hall conductivity $\chi_{xyy}$ at the bulk gap closing 
establish a one-to-one correspondence between nonequilibrium topological transitions and 
second-order transport. This sign flip is crucially a qualitative hallmark of a true topological 
transition: tuning the band tilt alone rescales $\chi_{xyy}$ monotonically without ever flipping 
its sign, while each Floquet gap closing does, making the nonlinear Hall response an unambiguous 
all-electrical probe of light-induced topology. Moreover, the magnitude and the dispersive 
lineshape of $\chi_{xyy}$ are tunable by drive strength, perpendicular electric field, Fermi 
energy, and temperature, thus offering several independent experimental knobs. These results 
establish that Floquet-driven \(1T'\)-MoS\(_2\) is a platform for nonlinear transport where a 
direct electrical fingerprint of nonequilibrium topological phases can be obtained without 
edge-state spectroscopy or magnetic probes. The predicted sign reversals of $\chi_{xyy}$ could 
be observed experimentally in second-harmonic Hall measurements as a function of drive 
strength~\cite{ma2019observation, kang2019}. Gate-tunable Fermi energy control offers an 
additional knob to map the dispersive dipole lineshape~\cite{PhysRevLett.121.246403,xu2014high}.

Finally, our study implies that nonlinear transport can provide a universal electrical method for finding nonequilibrium topological phases in driven quantum materials, complementing existing spectroscopic and edge-state measurements.

\bibliographystyle{apsrev4-2}
\bibliography{references}

\end{document}